\documentclass[10pt,a4paper]{article}
\usepackage[utf8]{inputenc}
\usepackage[english]{babel}
\usepackage{amsmath}
\usepackage{amsfonts}
\usepackage{amssymb}
\usepackage{graphicx}
\usepackage{lmodern}
\usepackage[left=2cm,right=2cm,top=2cm,bottom=2cm]{geometry}
\date{}
\begin{document}

\title{Thermodynamics of interacting tachyonic scalar field}
\author {S. D. Pathak$^{a}$\footnote{shankar@sdu.edu.cn}, M. M. Verma$^{b}$\footnote{sunilmmv@yahoo.com} and  Shiyuan Li$^a$\footnote{lishy@sdu.edu.cn}
        \\
        (a) School of Physics, Shandong University, Jinan 250100, China. \\
        (b) Department of Physics, University of Lucknow, Lucknow 274505 India. \\ }

\maketitle

\begin{abstract}
In this paper we discuss the laws of thermodynamics for interacting tachyonic scalar field. The components of the tachyonic scalar field in the universe are taken to exist in the state of non-equilibrium initially, but due to interaction they undergo a transition towards the equilibrium state. We show that the zeroth law of thermodynamics demands interaction among the components of cosmic field. The second law of thermodynamics is governing dynamics in transfer of energy among the three components of the proposed field with local violation of conservation of energy for individual components.
.
\end{abstract}

\section{Introduction}
The accelerated expansion of the universe reveal by a large number of cosmological observations \cite{i1} and  can be understood by introduction of repulsive gravity, although the other alternatives \cite{i6} may also explain the accelerated expansion. The repulsive gravity demands some different types of entities which have negative pressure. Considering its repulsive character such type of entity  is called dark energy.

A class of scalar fields is one of the promising candidates of dark energy. Among itself tachyonic scalar field appearing in the context of string theory \cite{i10} is logically more appealing than its counterpart quintessence due to its relativistic Lagrangian as analogue of particles. Cosmological relevance of this field has been studied by several authors during last few decades \cite{i11}.  Action and Lagrangian of the cosmological tachyonic scalar field is given by
 \begin{eqnarray}\mathcal{A}=\int d^{4}x \sqrt{-g} \left(\frac{R}{16\pi G}- V(\phi)\sqrt{1-\partial^{i}\phi\partial_{i}\phi}\right)\label{k1}\end{eqnarray}
 with Lagrangian $L=-V(\phi)\sqrt{1-\partial_{i}\phi\partial^{i}\phi}$ and the equation of motion of the field is found by varying action as
\begin{equation}\label{chap3.24}
  \frac{\ddot{\phi}}{\dot{\phi}}+ \frac{(1-\dot{\phi}^{2})V'(\phi)}{\dot{\phi}V(\phi)} + 3H(1-\dot{\phi}^{2})  = 0
\end{equation}
The stress energy tensor for this Lagrangian
\begin{eqnarray} T^{ik}=\frac{\partial{L}}{\partial{(\partial_{i}\phi)}}\partial^{k}\phi-g^{ik}L\label{n2}\end{eqnarray}
  This gives energy density and  pressure for spatially homogeneous field as \begin{eqnarray} \rho=\frac{V(\phi)}{\sqrt{1-\dot{\phi}^{2}}}; \qquad P=-V(\phi)\sqrt{1-\dot{\phi}^{2}}\label{n3}\end{eqnarray} respectively.

\section{Components of Field}

We assume  that radiation with equation of state $w_{r}=1/3$  also exists as  one inherent  component of  same tachyonic scalar field. Due to some physical
mechanism not known in detail at present to us, the cosmic tachyonic scalar field may be decomposed into several components with the assumption that the field is spatially homogeneous.  we can write the expressions for energy density and pressure as
\begin{eqnarray}P=-\frac{V(\phi)}{\sqrt{1-\dot{\phi}^{2}}} + \frac{\dot{\phi}^{2}V(\phi)}{\sqrt{1-\dot{\phi}^{2}}} + 0 \label{n8}\end{eqnarray}

\begin{eqnarray}\rho=\frac{V(\phi)}{\sqrt{1-\dot{\phi}^{2}}} + \frac{3\dot{\phi}^{2}V(\phi)}{\sqrt{1-\dot{\phi}^{2}}} - \frac{3\dot{\phi}^{2}V(\phi)}{\sqrt{1-\dot{\phi}^{2}}}.\label{n9}\end{eqnarray}
From (\ref{n8}) and  (\ref{n9}) it is seen  that when we include  radiation in tachyonic scalar field then one new exotic component also appears (say,  exotic matter since its energy density is negative) with zero pressure. Thus,  the tachyonic scalar field resolves into  three components say $a$, $b$ and $c$. The pressure and energy density  of $a$  is given as \[P_{a}=-\frac{V(\phi)}{\sqrt{1-\dot{\phi}^{2}}}, \quad \rho_{a}=\frac{V(\phi)}{\sqrt{1-\dot{\phi}^{2}}} \Rightarrow w_{a} = -1 = w_{\lambda}.\] This is nothing but the  `true' cosmological constant because of its equation of state  being $ w_{\lambda}=-1$.

The second component with   \[P_{b}=\frac{\dot{\phi}^{2}V(\phi)}{\sqrt{1-\dot{\phi}^{2}}},  \quad \rho_{b}=\frac{3\dot{\phi}^{2}V(\phi)}{\sqrt{1-\dot{\phi}^{2}}} \Rightarrow w_{b} = 1/3 \] can be identified as  radiation with $w_{r}=1/3$.  The  last component is characterised by  \[P_{c}=0,\quad \rho_{c}= -\frac{3\dot{\phi}^{2}V(\phi)}{\sqrt{1-\dot{\phi}^{2}}} \Rightarrow w_{c}=0.\] This component  mimics dust matter but has  negative energy density. The exotic matter may  include  the Dirac fermions as well as  the Majorana fermions whence the negative energy states turn into the positive energy states\cite{a5,b1}.
In our earlier work \cite{q1} we  allowed a  small time dependent perturbation in  the equation of state(EoS) of the cosmological constant with $\bar{w}_{\lambda}=-1+\varepsilon(t)$. Thus, with  the perturbed EoS, the true cosmological constant becomes  a shifted cosmological parameter.  This has a bearing upon the  EoS of radiation and exotic matter, both. Therefore,  these two entities  turn into  shifted radiation and shifted exotic matter respectively.  With fixed energy density of field components, the expressions for the  energy density and  pressure of each component are  given as below. For the shifted cosmological constant one has

\begin{eqnarray}
    \bar{\rho}_{\lambda}=\frac{V(\phi)}{\sqrt{1-\dot{\phi}^{2}}}; \qquad \bar{p}_{\lambda}=\frac{-V(\phi)}{\sqrt{1-\dot{\phi}^{2}}}+\frac{\varepsilon V(\phi)}{\sqrt{1-\dot{\phi}^{2}}}\label{n10}
\end{eqnarray}

and  $\bar{w}_{\lambda}=-1+\varepsilon(t)$.  For shifted radiation, we have
\begin{eqnarray}
    \bar{\rho}_{r}=\frac{3\dot{\phi}^{2}V(\phi)}{\sqrt{1-\dot{\phi}^{2}}}; \qquad \bar{p}_{r}=\frac{(1+3\varepsilon)\dot{\phi}^{2}V(\phi)}{\sqrt{1-\dot{\phi}^{2}}}\label{n12}
\end{eqnarray}
In presence of perturbation the zero pressure of exotic matter turns into negative non-zero pressure due to  shifted exotic matter which would  also accelerate the universe like dark energy. Thus,  the  energy density and pressure for shifted exotic matter are now,  respectively,   given as

\begin{eqnarray}
    \bar{\rho}_{m}=\frac{-3\dot{\phi}^{2}V(\phi)}{\sqrt{1-\dot{\phi}^{2}}}; \qquad
  \bar{p}_{m}=p_{\phi}-\bar{p}_{\lambda}-\bar{p}_{r}=\frac{-\varepsilon(1+3\dot{\phi}^{2})V(\phi)}{\sqrt{1-\dot{\phi}^{2}}}\label{n14}
\end{eqnarray}

with $\bar{w}_{m}=\frac{\varepsilon(1+3\dot{\phi}^{2})}{3\dot{\phi}^{2}}$.

\section{Thermodynamics Laws for interacting components}
Why  must the components of cosmic field interact? This is  one of the most interesting question  about interaction. Interaction might be justify by thermodynamics \cite{kp1}. As shown by obvious observations the  cosmic field must include at least three components representing matter, dark energy and radiation (many other components are also possible) and behave as an ensemble of three interacting thermodynamic systems. We apply the \textbf{Zeroth law of the thermodynamics} to these three systems called as shifted cosmological parameter (SCP), shifted radiation (SR) and shifted exotic matter (SEM) each.
\emph{\textrm{If SCP is in equilibrium with SR and SR is in equilibrium with SEM then SEM should be also in equilibrium with SCP}}.
This zeroth law demands interaction among cosmic field components whenever equilibrium gets perturbed for any reasons. If equilibrium is disturbed and the components are in non-equilibrium (thermal, mechanical or else), then  to re-attain the equilibrium the components must interact mutually.  If the components are in equilibrium then due to interaction perturbation in equilibrium state  reacts trying to restore its state  or achieve a new one  (Le Ch\^{a}telier-Braun principle) \cite{g1,g2}. If all components of tachyonic scalar field are in non-equilibrium,  then,  to achieve  an  equilibrium state they must fall into  mutual  interaction.  This motivation provides one justification to  study the interaction of these components. Here, we assuming that even though  the total energy of the  perturbed field (spatially homogeneous) is kept  conserved \textbf{(First law of thermodynamics)}  yet  during interaction it can  get  reasonably  violated for individual components.
We consider the three components of cosmic tachyonic scalar field the components are shifted cosmological parameter (SCP), shifted radiation (SR) and shifted exotic matter (SEM). The individual equations of energy conservation for SCP, SR and SEM  are respectively given as,

\begin{eqnarray}
\dot{\bar{\rho}}_{\lambda}+3H(1+\bar{w}_{\lambda})\bar{\rho}_{\lambda}=-Q_{1}\label{n16}
\end{eqnarray}
\begin{eqnarray}
\dot{\bar{\rho}}_{r}+3H(1+\bar{w}_{r})\bar{\rho}_{r}=Q_{2}\label{n17}
\end{eqnarray}
\begin{eqnarray}
\dot{\bar{\rho}}_{m}+3H(1+\bar{w}_{m})\bar{\rho}_{m}=Q_{1}-Q_{2}\label{n18}
\end{eqnarray}
where $Q_{1}$ and  $ Q_{2} $ are the interaction strengths and $H$ is Hubble parameter.
\textbf{Second Law of Thermodynamics} is the governing dynamics of interaction and  sign of $Q_{1}$ and $Q_{2}$ shows the direction of flow of energy density  during interaction among components. The positivity of the quantity $Q_1- Q_2 $ implies that $Q_1$ should be large and positive. For if $Q_1$ had been large and negative, then the second law of thermodynamics would have been violated and the SCP (as the dark energy candidate) would have dominated much earlier withholding the structure formation against the present observations. Also, $Q_2$ should be positive and small since if it is negative and large then conservation of energy of tachyonic field is violated.

\section{Interaction among the components}
The interacting dark energy models have been recently proposed by several authors \cite{j1}.
We study the interaction of these three components  assuming that even though  the total energy of the  perturbed field (spatially homogeneous) is kept  conserved , yet  during interaction it can  get  reasonably  violated for individual components.

In the above expressions (\ref{n16}),(\ref{n17}) and (\ref{n18}) the following  broad conditions must  govern the dynamics.

\textbf{Condition(I)}  $\mid Q_{1}\mid > \mid Q_{2}\mid$.  This corresponds  to  the  following cases,
\begin{description}
  \item[(i)] If $Q_{1} > 0$,  $Q_{2} > 0$  then the right hand side of (\ref{n16}) is negative while (\ref{n17}) and (\ref{n18}) are positive, respectively. This means that there is  energy transfer from shifted cosmological parameter to shifted radiation and shifted exotic matter,  respectively. Thermodynamics allows this kind of transfer of energy.
  \item[(ii)]  $Q_{1} < 0$,  $Q_{2} < 0$ implies  that there is an  energy transfer to shifted cosmological parameter from shifted radiation and shifted exotic matter.
\end{description}
\textbf{Condition(II)}  $ \mid Q_{2}\mid > \mid Q_{1}\mid$.
\begin{description}
  \item[(i)]  $Q_{2} > 0$,  $Q_{1} > 0$  would make  the right hand side of  (\ref{n17}) as  positive and  (\ref{n16}) and (\ref{n18}) as  negative. This shows  that there is an  energy transfer to shifted radiation from shifted cosmological parameter  and shifted exotic matter.  Thermodynamics again does  not allow this kind interaction.
  \item[(ii)]  $Q_{1} < 0$,  $Q_{2} < 0$ makes way for the   energy transfer from shifted radiation to  shifted exotic matter and  shifted cosmological parameter.
\end{description}

\textbf{Condition(III)} If  $Q_{2} = Q_{1}=Q$   then  we have the following possibility

\begin{description}
  \item[(i)] $Q>0$ leads to an  energy transfer to shifted radiation from shifted cosmological parameter, while the  shifted exotic matter remains free from interaction with its  energy density held conserved. This type of interaction holds compatibility with  the laws of thermodynamics.
  \item[(ii)] If  $Q<0$, energy would flow  from shifted radiation to shifted cosmological parameter, whereas the  shifted exotic matter does  not get involved  in interaction mechanism.  Thus,  the conservation of energy for shifted exotic matter holds good.

      \item[(iii)] As an  alternative, $Q=0$  would pull   the components of tachyonic scalar field  out of mutual interaction  like the standard $\Lambda$CDM model.
\end{description}
 The second case of condition (I) and condition (II) violate the laws of thermodynamics,  therefore,  we are  not interested  in these types of interactions. The interaction of type condition (iii) has been discussed for two components of  tachyonic scalar field in our earlier work \cite{a4}.  Due to the lack of information regarding the  exact nature of dark matter and dark energy (as the cosmological constant or else) we present the form of interaction term  heuristically  as  function of time rate of change in  energy densities as
\begin{eqnarray}Q_{1}=\alpha\dot{\bar{\rho_{\lambda}}}; \quad Q_{2}=\beta\dot{\bar{\rho_{r}}}\label{Q12}\end{eqnarray} where $\alpha, \beta$ are proportionality constant. While several authors have proposed different forms of $Q$ \cite{j1}.

From  (\ref{n16}), (\ref{n17}) and (\ref{n18}), for the specific dynamical form of interaction strengths (\ref{Q12}) one can found the functional form of energy density with redshift $z$ as,
\begin{eqnarray}
\bar{\rho}_{\lambda}=\bar{\rho}^{0}_{\lambda}x^{3\varepsilon/1+\alpha}\label{n22}
\end{eqnarray} where $\frac{a_{0}}{a}=1+z=x$
\begin{eqnarray}
\bar{\rho}_{r}=\bar{\rho}^{0}_{r}x^{4+3\varepsilon/1-\beta}\label{n23}
\end{eqnarray} and
\begin{eqnarray}
\bar{\rho}_{m}=\bar{\rho}^{0}_{m}x^{\eta}+\left(\frac{3\varepsilon\alpha\bar{\rho}^{0}_{\lambda}}{3\varepsilon-\eta-\eta\alpha}\right)[x^{3\varepsilon/1+\alpha}-x^{\eta}] -\left(\frac{\beta\bar{\rho}^{0}_{r}(4+3\varepsilon)}{4+3\varepsilon-\eta+\eta\beta}\right)[x^{4+3\varepsilon/1-\beta}-x^{\eta}]\label{n24}
\end{eqnarray}
where  $\eta$ assuming constant (with $\dot{\phi}^{2}\approx$ constant) is  defined as
\begin{eqnarray}
\eta=\frac{3\dot{\phi}^{2}(1+\varepsilon) + \varepsilon}{\dot{\phi}^{2}}.\label{n25}
\end{eqnarray}
Thus the cosmic expansion history of the universe is given by Hubble parameter with interaction as,

\begin{equation}\label{n26}
  H^{2}=\frac{\kappa^2}{3}[\bar{\rho}_{\lambda}+ \bar{\rho}_{r}+ \bar{\rho}_{m}]
\end{equation}
where $\kappa^{2}= 8\pi G$.

\section{Conclusion}
Having the motivation for the relativistic (tachyonic) scalar field, in contrast to quintessence, the single tachyonic scalar field which splits due to some unknown mechanism into three components (cosmological constant, radiation and dust matter). Due to consideration of radiation in this field the dust matter appears with negative energy. A small perturbation allowed in EoS of cosmological constant changes its status from a true cosmological constant to a shifted cosmological parameter (SCP). Similarly, status of radiation and dust matter changes to shifted radiation (SR) and shifted exotic matter (SEM). Thermodynamics laws (Zeroth, First and Second) might be responsible for interaction among components of single cosmic field. We consider the components of the field as thermodynamic systems and they interact to achieve a thermodynamical equilibrium. Particularly  Zeroth Law  invite interaction among components to maintain thermodynamical equilibrium or get new one, First Law demands the total energy of field stays conserved but the field components mutually interact with interaction strength parameter $Q$ resulting in local violation of energy conservation and Second Law decide the direction of flow of energy during interaction.  The entire evolution of the universe arises from this process of interaction.

\section{Acknowledgement}
We thank the members of the theoretical physics group of Shandong University at Jinan, China for helpful discussions. MMV also thanks Edward Rocky Kolb for the kind hospitality at Kavli Institute of Cosmological Physics, the University of Chicago, USA.

\end{document}